\documentclass[preprint,nobibnotes,twocolumn,10pt,byrevtex,prl]{revtex4}%
\usepackage{amsfonts}
\usepackage{amsmath}
\usepackage{amssymb}
\usepackage{graphicx}%
\providecommand{\U}[1]{\protect\rule{.1in}{.1in}}

\begin{document}
\preprint{UATP/1702}
\title{Comment on Ben-Amotz and Honig, "Average entropy dissipation in irreversible
mesoscopic processes," Phys. Rev. Lett. \textbf{96}, 020602 (2006) }
\author{P.D. Gujrati}
\email{pdg@uakron.edu}
\affiliation{Department of Physics, Department of Polymer Science, The University of Akron,
Akron, OH 44325}

\begin{abstract}
We point out that most of the classical thermodynamics results in the paper
have been known in the literature, see Kestin \cite{Kestin} and Woods
\cite{Woods}, for quite some time and are not new, contrary to what the
authors imply. As shown by Kestin, these results are valid for quasistatic
irreversible processes only and not for arbitrary irreversible processes as
suggested in the paper. Thus, the application to the Jarzynski process is limited.

\end{abstract}
\date{\today}
\maketitle

Ben-Amotz and Honig (BH) have suggested that some of their results in
\cite{Honig1} are "$\cdots$ either not previously recognized or not obtainable
nearly as transparently$\cdots$," but they do not clearly identify what they
are. They also do not clearly state whether their results are applicable
regardless of how fast the processes are carried out. Then there are some
results that may leave a reader somewhat confused or misinformed. This Comment
is intended to clarify these issues and to extend some results. The authors
begin by introducing the equality (see also \cite{Honig2,Honig})
\begin{equation}
\Delta S_{\text{Univ}}-\theta=0 \label{Entropy_Isolated}%
\end{equation}
for an isolated system (identified as "Univ") during a process $\mathcal{P}$
in terms of the entropy deficit function $\theta$, but then diminish its
importance by calling it to be "$\cdots$no more than a bookkeeping
convenience$\cdots$," which is unfortunate as this is one of the most
celebrated equation of nonequilibrium thermodynamics emphasizing the
irreversible entropy generation $\theta$, conventionally denoted by
$\Delta_{\text{i}}S$ \cite{Prigogine0,Prigogine}, due to \emph{dissipation
within the system}. Clausius \cite{Clausius} was the first to point to its
importance and identified it as the \emph{uncompensated transformation}
$N~$(see also Eq. (3.4.4) in \cite{Prigogine} and Eq. (3.18) in \cite{Eu}); it
was later elaborated by de Donder \cite{deDonder}. Treating the isolated
system as a composite system composed of the system (no sub) and the medium
(sub $0$), we have $\Delta S_{\text{Univ}}=\Delta S+\Delta S_{0}%
=\Delta_{\text{i}}S$, since by definition the medium has no irreversibility.
Therefore, $\Delta_{\text{i}}S$ is solely due to internal processes, quite
distinct from entropy exchange $\Delta_{\text{e}}S$ with the medium. The
\emph{clear separation} between $\Delta_{\text{e}}S$ and $\Delta_{\text{i}}S$
allowed Prigogine \cite{Prigogine} to write down the celebrated partition
$\Delta S=\Delta_{\text{e}}S+\Delta_{\text{i}}S$ (or $dS=d_{\text{e}%
}S+d_{\text{i}}S$ for infinitesimal changes). Use of calculus
(differentiation/integration), which is allowed due to the continuity of a
well-defined process $\mathcal{P}$ in the state space $\Omega$,\ gives
\begin{equation}
d_{\text{e}}S=dQ_{\text{irr}}/T_{0},\Delta_{\text{e}}S\doteq%
{\textstyle\int}
dQ_{\text{irr}}/T_{0} \label{Exchange_Entropy}%
\end{equation}
over $\mathcal{P}$; here, $T_{0}$ is the temperature of the medium. We hope
that the reader recognizes that the equality (\ref{Entropy_Isolated}) is more
than a bookkeeping; it has profound consequences.

There are two different ways the equality (\ref{Entropy_Isolated}) can be
interpreted. One is to consider it to express the entropy change between two
equilibrium (EQ) states after removing some constraints. The other
interpretation is to consider it to express the entropy change of an isolated
system as it relaxes towards equilibrium \cite[see Sec. 8]{Landau} and clearly
shows the internal nature of $\Delta_{\text{i}}S$. While BH restrict
themselves to processes between two EQ states, using the second interpretation
provides a deeper understanding of Clausius's uncompensated transformation in
our opinion. In general, $\Delta_{\text{i}}S$ consists of two independent
additive contributions $\Delta_{\text{i}}S^{\text{(Q)}}$ and $\Delta
_{\text{i}}S^{\text{(W)}}$ due to heat and work, respectively, where we mean
all sorts of work for the latter such as the pressure-volume work and the
chemical work. It should be clear that $\Delta_{\text{i}}S$, \textit{i.e.,
}$\theta$ in Eq. (\ref{Entropy_Isolated}) depends on the process $\mathcal{P}%
$. In an infinitesimal process $d\mathcal{P}$, its endpoints are
infinitesimally close in the state space to allow using differentiations
$dS,dQ$, etc.

The central concept BH present is a scheme \cite{Honig,Honig2} to introduce
another nonnegative deficit function $dW_{\text{diss}}\doteq dW_{\text{irr}%
}-dW_{\text{rev}}$; it requires comparing an irreversible process
$d\mathcal{P}_{\text{irr}}$ between two EQ states \textsf{A }and\textsf{ B
}with a reversible process $d\mathcal{P}_{\text{rev}}$ between the same two
states. It appears from \cite{Honig1,Honig2,Honig} and several other papers by
these authors listed in \cite{Honig3} that a reader most probably will get the
impression that the scheme is a creation of Honig \cite{Honig}, which is
factually incorrect. The scheme is very old and has been discussed by many
authors. It suffices to cite two old textbooks by Woods \cite[see Secs. 3, 5
and 7 and substitute $W_{\text{diss}}$ for $d^{\prime}W_{\text{i}}$]{Woods}
and by Kestin \cite[see Sec. 13.7 and substitute $W_{\text{diss}}$ for
$W_{\text{loss}}$]{Kestin}, and a monograph by Eu \cite[Secs. 4.2 and 4.3]{Eu}
where the same approach has been described in detail. Furthermore by using
$dQ_{\text{irr}}=T_{0}d_{\text{e}}S$ from Eq. (\ref{Exchange_Entropy}), we see
that Eq. (13.47) in \cite{Kestin} or Eq. (7.7) in \cite{Woods} is nothing but
the second of Eq. (5) in \cite{Honig1}. Its derivation by Kestin or by Woods
is just as simple if not more in my opinion. Similarly, Eq. (2) in
\cite{Honig1} appears as Eq. (13.50) in \cite{Kestin}; even the notation
$d\theta$ for the deficit function is also the same. The nonnegativity of
$(T_{0}-T)dS$ or $(T_{0}-T)d_{\text{e}}S$ discussed by BH follows trivially
from the second law conditions in Eq. (7.8) in \cite{Woods}, again using
$dQ_{\text{irr}}=T_{0}d_{\text{e}}S$. Therefore, it is factually incorrect for
BH to state that the results of classical thermodynamics in \cite{Honig1} that
follow from the first and second laws `$\cdots$\emph{were not previously
recognized or not obtainable nearly as transparently}$\cdots$.' It is worth
pointing out that BH are aware of Kestin \cite{Kestin} as they cite it in
\cite{Honig2}.

Let $\Omega(\mathbf{X})$ denote the EQ state space spanned by the set of $n$
observables $\mathbf{X}=\left\{  N,E,V,\cdots\right\}  $; it contains all EQ
states such as \textsf{A }and\textsf{ B }of the system. All states belonging
to $\mathcal{P}_{\text{rev}}$ or $d\mathcal{P}_{\text{rev}}$ between \textsf{A
}and \textsf{B} form a \emph{continuous} path in $\Omega(\mathbf{X})$ and the
derivative $1/T\doteq\partial S(\mathbf{X})/\partial E$ defines the
temperature. In contrast, nonequilibrium (NEQ) states are usually specified in
an \emph{extended} state space $\Omega(\mathbf{Z}_{m}),\mathbf{Z}%
_{m}\mathbf{=(X,}\boldsymbol{\xi}_{m}\mathbf{)}$, involving a set
$\boldsymbol{\xi}_{m}$ of independent internal variables
\cite{Maugin,Langer,Gujrati-II}, their number $m$ depending on how far the NEQ
state is from the EQ state, \textit{i.e., }how fast the process is carried
out: faster a process, larger is $m$. As the process becomes slower, some of
the internal variables no longer remain independent of $\mathbf{X}$ and are
not required so that $m$ decreases. The system's NEQ entropy $S$ is a state
function of $\mathbf{Z}_{m}$ and the system's temperature is the derivative
$1/T_{m}\doteq\partial S(\mathbf{Z}_{m})/\partial E$. As expected,
$\mathbf{\xi}_{m}$ also contributes to $\Delta_{\text{i}}S$. But BH only use
$\mathbf{X}$ and no $\boldsymbol{\xi}_{m}$.

Kestin takes pains to clarify that his Eq. (13.50) is only applicable to an
infinitesimal \emph{quasistatic }(qs)\emph{ irreversible} \emph{process}
$d\mathcal{P}_{\text{irr}}^{\text{qs}}$ for which the departures from
equilibrium are not large \cite[Sec. 4.6, in particular p. 133]{Kestin}. By
this, Kestin means as do Landau and Lifshitz \cite[see Sec. 13]{Landau} that
the NEQ states are still specified by $\mathbf{X}$ and
differentiation/integration remains meaningful; it is easily seen
\cite{Gujrati-I} that one still obtains nonzero $\Delta_{\text{i}}S$. This is
also the picture BH use in \cite{Honig2}. In contrast, BH claim in
\cite{Honig1} that their scheme is applicable to a $d\mathcal{P}_{\text{irr}}$
without any restriction on its speed so that it can be too far away from its
reversible analog $d\mathcal{P}_{\text{rev}}$. As a consequence, they state
that `$\cdots$the intermediate states of an irreversible process need not be
characterized by a well-defined temperature$\cdots$,' whereas
Kestin's\ qs-requirement does not have such a drawback. All it means is that
$T_{m}$ \emph{cannot} be uniquely determined if $\boldsymbol{\xi}_{m}$ is not
known. However, BH's claim creates a logical problem as $dQ_{\text{irr}}$
depends on the temperatures of the system and the heat bath over the process
and may remain ambiguous or even undetermined for $m>0$ and so it could not be
\emph{determined}, i.e.,\emph{ measured}. Therefore, $dW_{\text{irr}%
}=dE-dQ_{\text{irr}}$ suffers from the same fate and Eqs. (3) and (5) would be
useless for $m>0$ as $dW_{\text{irr}}$ is \emph{undetermined}. To see this, we
divide $d\mathcal{P}_{\text{irr}}$ into a finite collection of still "smaller"
and nonoverlapping subprocesses $\{d\mathcal{P}_{\text{irr}}^{(k)}%
\},d\mathcal{P}_{\text{irr}}\equiv\cup_{k}d\mathcal{P}_{\text{irr}}^{(k)}$,
involving intermediate states. We denote their heat by $dQ_{\text{irr}}^{(k)}$
so that $dQ_{\text{irr}}=%
{\textstyle\sum\nolimits_{k}}
dQ_{\text{irr}}^{(k)}$. However, we cannot determine those $dQ_{\text{irr}%
}^{(k)}$'s for which the corresponding temperature $T_{m}^{(k)}$ is not known.
The problem becomes even more serious if we consider a finite process
$\mathcal{P}_{\text{irr}}$. In order to determine the cumulative heat $\Delta
Q_{\text{irr}},\theta$, etc. we need to `$\cdots$\emph{integrate}$\cdots
$\emph{ around the actual path of interest}' $\mathcal{P}_{\text{irr}}$ as BH
suggest \cite{Honig1,Honig2,Honig}. What is important is that we need to know
the temperature and other properties of the intermediate states of
$\mathcal{P}_{\text{irr}}$ \emph{uniquely}; but it may not be well defined if
$\boldsymbol{\xi}_{m}$ is not known. Moreover, if $\mathcal{P}_{\text{irr}}$
is fast enough, then no reversible path can be found for the elemental process
$d\mathcal{P}_{\text{irr}}$ for which at least one of the end states requires
some internal variables ($m>0$) for its unique specification. As a result, the
BH scheme and in particular Eqs. (3) and (5) would fail for some finite processes.

The situation can be salvaged provided we consider qs-irreversible processes
recommended by Kestin \cite[p. 133]{Kestin} and followed in \cite{Honig2} as
they result in\emph{ continuous paths }between \textsf{A }and \textsf{B} in
$\Omega(\mathbf{X})$. Thus, the BH scheme, although it appears sensible, seems
to be of limited use and cannot be applied to arbitrary irreversible processes
such as the Jarzynski process $\mathcal{P}_{\text{J}}$ \cite{Jarzynski}, which
has no restriction on how irreversible it could be so that it may have no
continuous paths between \textsf{A }and\textsf{ B} in $\Omega(\mathbf{X})$.
Despite this, it is said to determine the difference $\Delta A$ of the
equilibrium free energies between \textsf{A }and\textsf{ B} that is impossible
to measure as all processes are somewhat irreversible. The terminal states
\textsf{A }and\textsf{ B} of $\mathcal{P}_{\text{J}}$ have the same
temperature $T^{0}$, but the temperature during $\mathcal{P}_{\text{J}}$ may
be very different from it or may not even be defined.

\end{document}